\documentclass[amsmath,amssymb,superscriptaddress,aps,showpacs,10pt,twocolumn]{revtex4-1}

\usepackage{amsfonts}
\usepackage{graphicx}
\usepackage{soul}

\newcommand{\kb}[1]{\textrm{\tiny{#1}}}

\newcommand{\kd}[1]{\mathbf{#1}}

\begin{document}

\title{Electrically Dressed Ultralong-Range Polar Rydberg Molecules}

\date{\today}
\author{Markus Kurz}
\affiliation{Zentrum f\"ur optische Quantentechnologien, Luruper Chaussee 149, 22761 Hamburg, Universit\"at Hamburg, Germany}
\author{Peter Schmelcher}
\affiliation{Zentrum f\"ur optische Quantentechnologien, Luruper Chaussee 149, 22761 Hamburg, Universit\"at Hamburg, Germany}
\affiliation{The Hamburg Centre for Ultrafast Imaging, Luruper Chaussee 149, 22761 Hamburg, Universit\"at Hamburg, Germany}
\begin{abstract}
We investigate the impact of an electric field on the structure of ultralong-range polar diatomic Rubidium Rydberg
molecules. Both the $s$-wave and $p$-wave interactions of the Rydberg electron and the neutral ground state atom
are taken into account. In the presence of the electric field the angular degree of freedom between the electric field
and the internuclear axis acquires vibrational character and we encounter two-dimensional oscillatory
adiabatic potential energy surfaces with an antiparallel equilibrium configuration. The electric field allows
to shift the corresponding potential wells in such a manner that the importance of the $p$-wave interaction 
can be controlled and the individual wells are energetically lowered at different rates. As a consequence
the equilibrium configuration and corresponding energetically lowest well move to larger internuclear
distances for increasing field strength. For strong fields the admixture of non-polar molecular Rydberg
states leads to the possibility of exciting the large angular momentum polar states via two-photon processes
from the ground state of the atom. The resulting properties of the electric dipole moment and the vibrational
spectra are analyzed with varying field strength.
\end{abstract}
\maketitle
\section{Introduction}
Ultracold atomic and molecular few- and many-body systems offer a unique platform for a detailed
understanding and analysis of fundamental quantum properties.
The preparation and control of such systems in specific quantum states offer many opportunities for exploring
elementary quantum dynamical processes. Experimentally, one can control the external motion 
of the atoms by designing and switching between almost arbitrarily shaped traps \cite{Pethick08,Grimm00,Folman02},
and the strength of the interaction among the atoms can be tuned by magnetic or optical Feshbach resonances 
\cite{Koehler06,Bloch08,Chin10}. A striking new species are the weakly bound ultralong-range diatomic
molecules composed of a ground state and a Rydberg atom whose existence has been predicted theoretically more than a decade ago \cite{Greene00} and which have been discovered experimentally only recently \cite{Bendkowsky09}. 
The molecular Born-Oppenheimer potential energy curves, which are responsible for the atomic binding, show for these species a very unusual oscillatory behavior with many local minima. The latter can be understood intuitively and modeled correspondingly as the interaction of a neutral ground state atom with the Rydberg electron of the second atom. In a first approximation, the interaction between the two constituents is described by a $s$-wave scattering dominated Fermi-pseudopotential \cite{Fermi34,omont77}. The equilibrium distance for these molecular states is of the order of the size of the Rydberg atom and the vibrational binding energies are in the MHz to GHz regime for principal quantum numbers $n \approx 30-40$ depending on the type of states. More specifically, low-angular momentum non-polar states and large angular momentum polar states, so-called trilobites, have been predicted \cite{Greene00}.They possess electric dipole moments in the range of 1Debye (low-$\ell$)\cite{Pfaunat11} up to 1kDebye (high-$\ell$)\cite{Greene00} in the polar case. The large electric dipole moment of the latter makes them accessible for electric field manipulation which opens the interesting possibility for the external control
of molecular degrees of freedom. Beyond the $s$-wave interactions, $p$-wave scattering has been shown \cite{Hamilton02} to lead to a class of shape-resonance-induced long-range molecular Rydberg states.\\
The impact of magnetic fields on these ultralong-range molecules has been studied in \cite{Lesanovsky07} where it has
in particular been shown that the magnetic field provides an angular confinement turning a rotational degree
of freedom into a vibrational one and yields, with increasing strength, a monotonic lowering of the magnitude of the electric dipole moment. Polyatomic ultralong-range molecules formed of a Rydberg atom and several ground state perturber, such as collinear triatomic species, can be constructed by taking the diatomic wave function as a basic unit and constructing the corresponding symmetry-adapted orbitals \cite{Liu06}. Recently the formation of Rydberg trimers and excited dimers bound by internal quantum reflection \cite{Bendkowsky10} have been observed experimentally and analyzed in detail theoretically \cite{Bendkowsky10}.  
Moreover, it it has been shown how the electric field of a Rydberg electron can bind a polar molecule 
to form a giant ultralong-range stable triatomic molecule \cite{Rittenhouse10,Rittenhouse11,Mayle12} which can consequently be controlled by applying external electric fields. Combining electric and magnetic fields in a crossed field configuration the existence and properties of so-called giant dipole ultralong-range molecules have been shown \cite{Kurz12}. Opposite to the above-mentioned Rydberg molecules this species has no open radiative decay channels.

In spite of the diversity of works focusing on ultralong-range molecules an original investigation of the impact
of external electric fields specifically on the polar trilobite states is missing. Such an investigation is 
particularly desirable due to the strong sensitivity of these Rydberg molecules with respect to the external
field which provides a handle on the control of their properties on a single molecule basis but also for their
interactions in potential many-body systems. For these reasons we perform 
in this work a study of the impact of an electric field on the structure and dynamics of high-$\ell$ ultralong-range diatomic Rubidium molecules. We hereby proceed as follows. Section II provides a formulation of the problem presenting the working Hamiltonian and 
a discussion of the underlying interactions. Our analysis goes beyond the $s$-wave approximation and takes into account the next 
order $p$-wave term of the Fermi-pseudopotential. Section III and IV contain our methodology and a
discussion of the effects of the $p$-wave contribution, respectively. In section V we analyze the evolution of the topology of the
potential energy surfaces (PES) with changing electric field. The resulting PES show a strongly
oscillatory behavior with bound states in the MHz and GHz regime. With increasing field strength the diatomic
molecular equilibrium distance shifts substantially in a range of the order of thousand Bohr radii. We analyze the behavior
of the corresponding electric dipole moment thereby achieving molecular states with a dipole moment up to several kDebye.
Based on these properties and the  $s$-wave admixture via the external electric field
a preparation scheme for high-$\ell$ polar molecular electronic states via a two photon
excitation process is presented. Finally, we provide an analysis of the vibrational spectra which exhibit spacings
of the order of several MHz. 

\section{Molecular Hamiltonian and Interactions}
We consider a highly excited Rydberg atom interacting with a ground state neutral perturber atom 
(we will focus on the $^{87}$Rb atom here) in a static and homogeneous electric field. 
The Hamiltonian treating the Rb ionic core and the neutral perturber as point particles is given by
(if not stated otherwise, atomic units will be used throughout)
\begin{eqnarray}
H&=&\frac{\kd{P}^{2}}{M}+H_{\rm{el}} +V_{\textrm{n,e}}(\kd{r},\kd{R}),\label{ham}\\
H_{\rm{el}}&=&H_0 + \kd{E}\kd{r},\ \ H_0=\frac{\kd{p}^2}{2m_{\kb{e}}} + V_{l}(r), 
\end{eqnarray}
where $(M,\kd{P},\kd{R})$ denote the atomic $Rb$ mass and the relative momentum and position of the neutral perturber with respect
to the ionic core. $(m_{\rm{e}},\kd{p},\kd{r})$ indicate the corresponding quantities for the Rydberg electron. 
The electronic Hamiltonian $H_{\rm{el}}$ consists of the field-free Hamiltonian $H_0$ and the usual Stark term of an 
electron in a static external $\kd{E}$-field. 
$V_l(\kd{r})$ is the angular momentum-dependent one-body pseudopotential felt by the valence electron when interacting 
with the ionic core. For low-lying angular momentum states the electron penetrates the finite ionic $\text{Rb}^{+}$-core 
which leads to a $\ell$-dependence of the interaction potential $V_l(r)$ due to polarization and scattering effects 
\cite{Gallagher}. Throughout this work we choose the direction of the field to coincide with the z-axis of the 
coordinate system, i.e.\ $\kd{E}=E\kd{e}_{\rm{z}}$. Finally, the interatomic potential $V_{\rm{n},e}$ for the 
low-energy scattering between the Rydberg electron and the neutral perturber is described as a so-called Fermi-pseudopotential 
\begin{eqnarray}
V_{\textrm{n,e}}(\kd{r},\kd{R})&=&2\pi A_{s}[k(R)]\delta(\kd{r}-\kd{R})\\
&+&6\pi  A^{3}_{p}[k(R)]\overleftarrow{\nabla} \delta(\kd{r}-\kd{R}) \overrightarrow{\nabla}.\label{inter}
\end{eqnarray}
Here we consider the triplet ($S=1$) scattering of the electron from the spin-$\frac{1}{2}$ ground state alkali atom. 
Suppression of singlet scattering events can be achieved by an appropriate preparation of the initial atomic gas. 
In eq.\ (\ref{inter}) $A_{s}[k(R)]=-\tan(\delta_0(k))/k$ and $A^{3}_{p}[k]=-\tan(\delta_1(k))/k^{3}$ denote the energy-dependent 
triplet $s$- and $p$-wave scattering lengths, respectively, which are evaluated from the corresponding phase shifts 
$\delta_l(k),\ l=0,1$. The kinetic energy $E_{\rm{kin}}=k^2/2$ of the Rydberg electron at the collision point with the 
neutral perturber can be approximated according to $k^2 / 2=1/R-1/2n^2$. The behavior of the energy-dependent phase shifts 
$\delta_{l}$ as functions of the kinetic energy $E_{\rm{kin}}$ is shown in Fig.\ \ref{phase_shifts}. 
\section{Methodology}
In order to solve the eigenvalue problem associated with the Hamiltonian \eqref{ham} we adopt an adiabatic ansatz for 
the electronic and heavy particle dynamics. We write the total wave function as $\Psi(\kd{r},\kd{R})=\psi(\kd{r}; \kd{R} )\phi(\kd{R})$ and 
obtain within the adiabatic approximation
\begin{eqnarray}
[H_0 + \kd{E}\kd{r} +V_{\kb{n,e}}(\kd{r},\kd{R})]\psi_{i}(\kd{r};\kd{R})&=&\epsilon_{i}(\kd{R})\psi_{i}(\kd{r};\kd{R}), \label{hamelec}\\
(\frac{\kd{P}^{2}}{M}+\epsilon_{i}(\kd{R}))\phi_{ik}(\kd{R})&=&E_{ik}\phi_{ik}(\kd{R}),\label{hamrovi}
\end{eqnarray}
where $\psi_i$ describes the electronic molecular wave function in the presence of the neutral perturber for a given 
relative position $\kd{R}$ and $\phi_{ik}$ determines the rovibrational state of the perturber. To calculate the potential 
energy surface $\epsilon_{i}(\kd{R})$ we expand $\psi(\kd{r};\kd{R})$ in the eigenbasis of $H_0$, 
i.e.\ $\psi_i(\kd{r};\kd{R})=\sum_{nlm} C^{(i)}_{nlm}(\kd{R})\chi_{nlm}(\kd{r})$ with 
$H_0\chi_{nlm}(\kd{r})=\varepsilon_{nl}\chi_{nlm}(\kd{r}),\ \chi_{nlm}(\kd{r})=R_{nl}(r)Y_{lm}(\theta,\phi)$. 
For $l \ge l_{\rm{min}}=3$ we neglect all quantum defects, i.e.\ $H_0$ is identical to the hydrogen problem. 
Finally, we have to solve the following eigenvalue problem
\begin{eqnarray}
&\ &(\varepsilon_{nl}-\epsilon(\kd{R}))C_{nlm}+\sum_{n'l'm'}C_{n'l'm'}(E\langle nlm |z|n'l'm \rangle\delta_{mm'} \nonumber\\
&+&\langle nlm |V_{\textrm{n,e}}(\kd{r},\kd{R})|n'l'm' \rangle) = 0,
\end{eqnarray}
for which we use standard numerical techniques for the diagonalization of hermitian matrices. Throughout this work 
we mainly focus on the high-$\ell$ $n=35$ manifold which provides the trilobite states in case of zero electric field 
\cite{Greene00}. To ensure convergence we vary the number of basis states finally achieving a relative accuracy 
of $10^{-3}$ for the energy. For the $n=35$ trilobite manifold we used, in addition to the degenerate 
$n=35,\ l \ge 3$ manifold, a basis set that includes the $38s,\ 37d,\ 36p$ quantum defect split states due to their
energetically closeness. This basis set contains $1225$ states in total. \\
From eqs.\ (\ref{hamelec}) and (\ref{hamrovi}) we already deduce some symmetry properties of the states $\psi,\ \phi$ 
and the energies $\epsilon$. If $P_{\kd{r},\kd{R},\kd{E}}$ denotes the generalized parity operator that transforms 
$(\kd{r},\kd{R},\kd{E})\rightarrow(-\kd{r},-\kd{R},-\kd{E})$ we have 
$[H,P_{\kd{r},\kd{R},\kd{E}}]=[V_{\rm{n,e}}(\kd{r},\kd{R}),P_{\kd{r},\kd{R},\kd{E}}]=0$. This means that the states 
$\Psi,\ \psi$ and $\phi$  are parity (anti)symmetric and the PES fulfill $\epsilon(\kd{R};\kd{E})=\epsilon(-\kd{R};-\kd{E})$. 
In addition, the PES possess an azimuthal symmetry, e.g.\ the vector defining the internuclear axis can, without loss of 
generality, be chosen to lie in the $x-z$-plane. In the absence of any field, the PES depend exclusively on the 
internuclear distance $R$. However, if a field is present, the PES are cylindrical symmetric, which means they also depend on the 
angle of inclination $\theta$ between the field vector and the internuclear axis, e.\ g.\ $\epsilon(\kd{R})=\epsilon(R,\theta)$.
\section{Discussion of the p-wave contribution}
In several previous works the interaction between the Rydberg electron and the neutral perturber has been
modeled to consist exclusively of a $s$-wave scattering potential \cite{Greene00, Lesanovsky07}. In this case,
a single potential curve splits away from the $n^2-l^2_{\kb{min}}$ degenerate high-$l$ manifold forming a strongly oscillating
Born-Oppenheimer potential energy surface with a depth of around $-3.5 \times 10^5\ \text{GHz}/n^3$ providing rovibrational states
with a level spacing of approximately $100$MHz and a permanent electric dipole moment of $1$kDebye. However, it is an
important fact that the $p$-wave scattering length $A_{\rm{p}}$ possesses a resonance at $E_{\rm{kin}}=24.7$meV. This corresponds
to a radial distance of $756 a_0$ for a Rydberg electron in a $n=35$ state. The effect of the $p$-wave contribution and its
resonant behavior on adiabatic potential curves has been studied in detail in several works, see
\cite{Fabrikant02,Hamilton02}. It has been shown that due to this interaction additional PES split away from the degenerate
manifold leading to avoided crossings with the pure $s$-wave dominated potential curve and even energetically lower 
$s,\ p$ and $d$-states. These avoided crossings cause a dramatic change in the topology of the PES leading to novel molecular dynamics.\\
\begin{figure}[hbt]
\begin{minipage}{0.5\textwidth}
\includegraphics[width=1.0\linewidth]{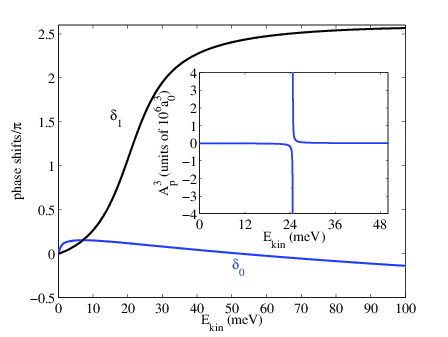}
\caption{(Color online) Energy dependent triplet phase shifts $\delta_1$ and $\delta_0$ for $e^{-}-^{87}$Rb($5s$) scattering. For $E_{\rm{kin}}=24.7$meV the phase shift $\delta_1=\pi/2$, i.e.\ the (cubed) energy dependent $p$-wave scattering length $A^{3}_{p}(k)=-\tan(\delta_1(k))/k^3$ possesses a resonance at this energy. This can be clearly seen in the inset.} 
\label{phase_shifts} 
\end{minipage}
\end{figure}
According to \cite{Kazansky92,Fabrikant02}, for large radial distances the overall behavior of $s$ and $p$-wave scattering
dominated PES can be well described by the Borodin and Kazansky model \cite{Kazansky92}
\begin{eqnarray}
\hspace{-0.25cm}\epsilon_{nl}(R)=-\frac{1}{2(n-\delta_l(k(R))/\pi)^2}\approx-\frac{1}{2n^2}-\frac{\delta_{l}(k(R))}{\pi n^3}.\label{borodin} 
\end{eqnarray}
The energy curves are roughly determined by the corresponding phase shifts for the electron-neutral atom scattering, 
although the individual oscillations necessary for the existence of stable molecular states cannot be described by the simple equation
\eqref{borodin}. It can be shown that the positions of the crossing point of both PES $R^{(n)}_{\rm{c}}$ and the position
of the minimum of the $s$-wave dominated PES $R^{(n)}_{\rm{m}}$ obey the ratio
\begin{eqnarray}
\frac{R^{(n)}_{\rm{c}}}{R^{(n)}_{\rm{m}}}=\frac{E^{(\rm{m})}_{\kb{kin}}+1/2n^2}{E^{(\rm{c})}_{\kb{kin}}+1/2n^2} 
\end{eqnarray}
with $E^{(\rm{c})}_{\rm{kin}}=7.1104$meV and $E^{(\rm{m})}_{\rm{kin}}=7.1102$meV being the corresponding electronic
kinetic energies (see Fig.\ \ref{phase_shifts}). Since $0.999973 \le R^{(n)}_{\rm{c}}/R^{(n)}_{\rm{m}} < 1$ it is in particular not
possible to substantially shift the avoided crossing between these
two PES by increasing the principal quantum number $n$ in the absence of the external field. This means
that the $p$-wave contribution is crucial for the overall topology of the PES.
\section{Potential energy surfaces}
The mechanism underlying the oscillating behaviour of the potential energy surfaces for ultralong-range molecules composed of a Rydberg atom plus neutral ground state atom is the following. The neutral atom is to a good approximation point-like and its interaction with the Rydberg atom probes the highly excited electronic wave function locally in space, meaning that the highly oscillatory character of the Rydberg wave function is mapped onto the potential energy surface. This holds both for the absence and presence of an electric field. 

\begin{figure}[hbt]
\begin{minipage}{0.5\textwidth}
\includegraphics[width=1.0\textwidth]{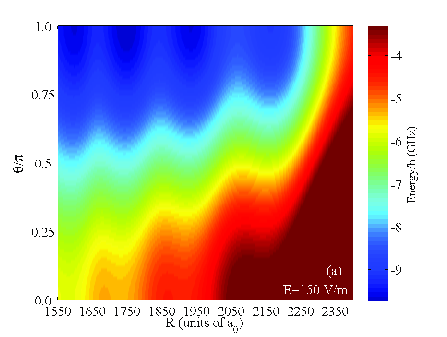}
\includegraphics[width=1.0\textwidth]{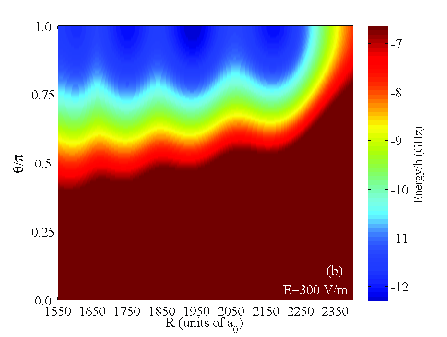}
\caption{(Color online) Two-dimensional potential energy surfaces for the electrically dressed polar trilobite
states for $E=150\frac{V}{m}$ (a) and $300\frac{V}{m}$ (b). We observe a potential minimum at $\theta=\pi$. An increase of the electric field goes along with a stronger confinement of the angular motion and an increase of the diatomic equilibrium distance $R_{\rm{eq}}$. Thus, the electric field stabilizes the $s$-wave dominated molecular states.} 
\label{pes}
\end{minipage}
\end{figure}

\begin{figure}[hbt]
\begin{minipage}{0.5\textwidth}
\includegraphics[width=1.0\textwidth]{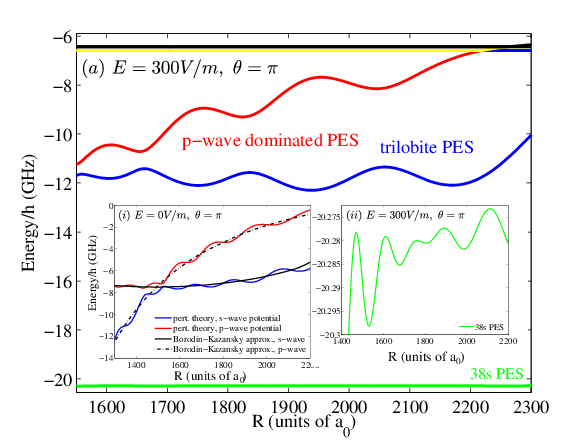}
\includegraphics[width=1.0\textwidth]{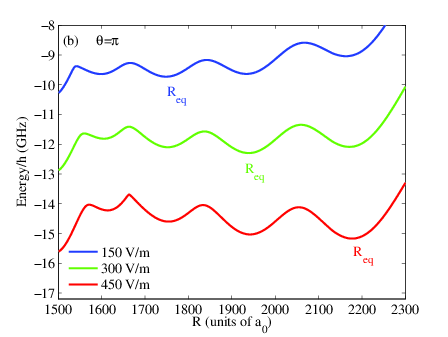}
\caption{(Color online) (a) Intersections through the two-dimensional PES for $\theta=\pi$ for the $9$th-$15$th excitation for $E=300 \frac{V}{m}$. For the two lowest PES arising from the high-$\ell$ degenerate manifold a strongly oscillatory behavior is visible. The region of avoided crossing is clearly visible at $R \sim 1400-1500a_0$. In addition, we present
the potential curves provided by the Borodin-Kazansky model given by eq.\ (\ref{borodin}).
The inset (i) shows the field-free trilobite and first $p$-wave PES. The inset (ii) shows the $38s$ split PES which is oscillatory in the MHz regime and hardly affected by the electric field. (b) Same as in (a) but with varying $E=150,\ 300$ and $450\frac{V}{m}$. The diatomic equilibrium distance $R_{\rm{eq}}$ is moving away from the region of the avoided crossing at $R=1500a_0-1700a_0$.}
\label{pes_curves}
\end{minipage}
\end{figure}

For the non-polar low-angular momentum states which are states that are splitted from the (degenerate) hydrogenic
manifold by a sizable quantum defect \cite{Greene00} the oscillations of the potential energy typically amount to                                                                                                                                                                                                                                                                                                                                                                                                                                                                                                                   
many MHz. The $38s$ state shown in Fig. (\ref{pes_curves}a) on a GHz scale and in the corresponding inset (ii) enlarged on a MHz scale, is such a non-polar state. Its oscillatory behaviour is weaker than the oscillatory behaviour of the polar trilobite or p-wave states which are in the GHz regime, see also Fig.\ (\ref{pes_curves}a) (many bound vibrational states exist in both cases). This is due to the energetical lowering in the framework of the mixing of the many high-angular momentum states available to form the trilobite state. Indeed, it has been shown in ref.(\cite{Greene00}) that for a pure $s$-wave interaction potential of the Rydberg electron and the neutral perturber the trilobite PES is given by
\begin{eqnarray}
\epsilon(R)=-\frac{1}{2n^2}+ \frac{1}{2} A_{s}[k(R)]\sum\limits^{n-1}_{l=l_{\rm{min}}}(2l+1)R^{2}_{nl}(R),\label{swave}
\end{eqnarray}
where $R_{nl}$ denote the hydrogenic radial functions \cite{Greene00}. Let us now focus on the case of the presence of an electric field specifically on the regime $E=0 - 650 \frac{V}{m}$. The dissociation limits
correspond to the atomic states Rb(5$s$)+Rb($n=35,l \ge 3$) where $l$ is used as a label in the presence of
the electric field. In Fig.\ \ref{pes} we present the PES for the electrically dressed polar trilobite
state for $E=150\frac{V}{m}$ and $300\frac{V}{m}$  as a function $\theta$ and $R$. As mentioned in section III in the absence of an electric field the potential curves are independent of $\theta$.  
For a finite field strength, this spherical symmetry is broken which is clearly seen in Fig.\ \ref{pes}. For all field strengths the potential minimum is taken for the antiparallel field configuration $\theta=\pi$. This is reasonable because the external electric fields forces the electron density to align in its direction which leads to a higher density in the negative $z$-direction. The electric field therefore turns a rotational degree of freedom $\theta$ to a vibrational one. As the field strength increases a stronger confinement of the angular motion is achieved and the corresponding equilibrium distance $R_{\rm{eq}}$ increases substantially.\\
In Fig.\ (\ref{pes_curves}a) we show intersections through the PES for the $9$th-$15$th excitation for a field strength of $300\frac{V}{m}$ for $\theta=\pi$. In addition we present two insets. Inset (i) in this figure shows the high-$\ell$ field-free potential curves. Due to the $p$-wave interaction a single potential curves splits away from the degenerate manifold causing an avoided crossing in the region of $R=1400a_0-1500a_0$. In addition, the inset also shows the potential curves provided by the Borodin-Kazansky approximation (see eq.\ (\ref{borodin})). In the main figure the lowest potential curve is the one belonging to the $38s$ quantum defect split state. This state possesses a very weak oscillatory behavior in the MHz regime. This can be clearly seen in inset (ii) of Fig.\ (\ref{pes_curves}a). In general, this state is much less affected by the electric field compared to the PES arising from the zero field high-$\ell$ degenerate manifold. This is reasonable since the atomic $38$s state does not possess a substantial electric dipole moment in the presence of the field. Therefore its potential curve hardly shifts with increasing electric field strength from its field-free value of $-20.284$GHz.\\
However, the potential curves arising from the $n=35, l \ge 3$ manifold show a strong dependence on the electric field. Analogously to the field free case (see inset (i) in Fig.\ (\ref{pes_curves}a)) we obtain potential curves with a strongly oscillatory structure in the many hundred MHz to GHz regime. It is important to note that the $p$-wave interaction dominated PES (red curve in Fig.\ (\ref{pes_curves}a)) and the $s$-wave interaction dominated PES (blue curve in Fig.\ (\ref{pes_curves}a)) exhibit an avoided crossing which is crucial for the stability of the corresponding vibrational states.           
For zero field this avoided crossing happens to be comparatively close to the global minimum of the corresponding 'trilobite' PES (see (i) in Fig.\ (\ref{pes_curves}a)) and might influence extended vibrational states.
For increasing field strength however this avoided crossing is
increasingly separated from the global equilibrium distance, as we shall discuss in more detail below.\\
In Fig.\ (\ref{pes_curves}b) we show intersections for $\theta=\pi$  through the PES of the electrically dressed polar states for different field strengths $E=150,\ 300,$ and $450 \frac{V}{m}$.
We observe how the potential curve is globally shifted with increasing electric field strength.
For $E \ge 700 \frac{V}{m}$ (not shown in Figure) this trilobite PES experiences avoided crossings with the potential curve                                                                                                                                                                                                                                                                                                                                                                                                                                                                                                                                                                                                                                                                                                                                                                                             
belonging to the $38s$ state. In addition,                                                                                                                                                                                                                                                                                                                                                         we see that the overall topology of the PES do not change with varying $E$-field. In particular, the number of minima and their positions remain approximately constant with increasing field strength. However, the diatomic equilibrium distance $R_{\rm{eq}}$ (which is the global minimum in the range $1550a_0\le R \le 2450a_0$) changes strongly as $E$ varies. This means that the electric field causes a 'spatial weight' to the PES. In contrast, the region of avoided crossing of the $s$-/$p$-wave dominated potential curves is hardly affected by the applied electric field and remains in the interval $1500a_0$ to $1700a_0$. This means that low-lying vibrational molecular states in the well around the global minimum are shifted away from the region of the avoided crossing. As a consequence the importance of the $p$-wave interaction is decreased significantly for higher field strength and the PES are determined mainly
by the $s$-wave interaction. To be more specific we show the dependence of $R_{\rm{eq}}$ as a function of $E$ in the inset in Fig.\ \ref{elec}. We see a plateau-like structure with steps at the field strengths $100,\ 200,\ 385 \frac{V}{m}$ where the value of $R_{\rm{eq}}$ sharply changes. This structure simply reflects the depicted effect of the electric field on the PES, i.e. by varying the electric field one changes the energetically position of the different potential wells in the oscillating PES, which leads to abrupt changes of the global equilibrium position $R_{\rm{eq}}$. 
Figs.\ (\ref{pes_curves}a,b) also demonstrate that with increasing field strength the avoided crossing between 
and the $s$-wave interaction dominated states remains (approximately) localized in coordinate space whereas the energetically low-lying
potential wells with bound vibrational states and in particular the one belonging to the global equilibrium position are lowering in energy and
are consequently well-separated from this avoided crossing. In conclusion, the electric field represents an excellent tool to control the energetic
positions and depths of the individual wells and to avoid destabilizing avoided crossing.

\section{Electric dipole moment}
\begin{figure}[hbt]
\begin{minipage}{0.5\textwidth}
\includegraphics[width=1.0\linewidth]{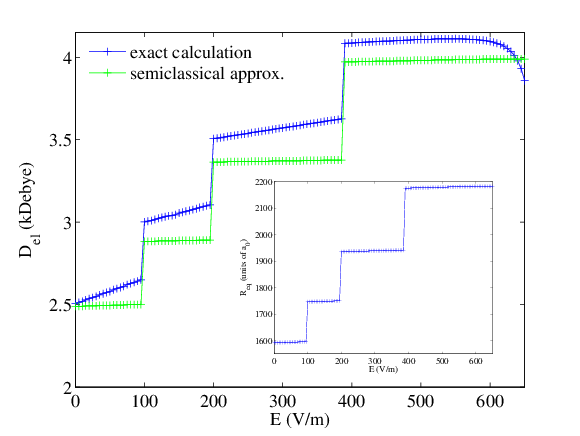}
\caption{(Color online) The electric dipole moment as a function of the electric field $E$ (blue points). For comparison we show a semi-classical prediction (green points). The inset shows the behavior of the equilibrium distance $R_{\rm{eq}}$ with varying electric field strength.}
\label{elec}
\end{minipage}
\end{figure}

\begin{figure}[hbt]
\begin{minipage}{0.5\textwidth}
\includegraphics[width=1.0\textwidth]{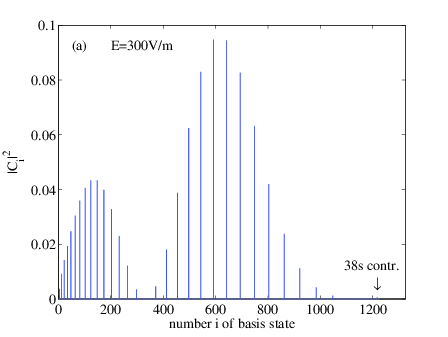}
\includegraphics[width=1.0\textwidth]{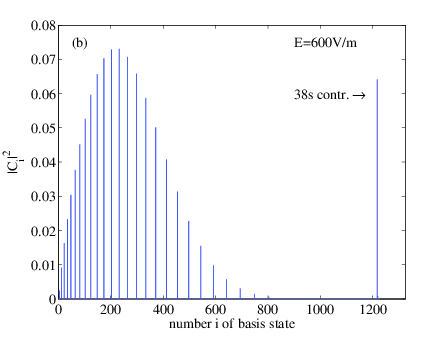}
\caption{(Color online) Spectrum of coefficients of the electronic eigenvector $\psi(\kd{r};\kd{R}_{\rm{eq}},\kd{E} )$ at $E=300 \frac{V}{m}$
(a) and $600 \frac{V}{m}$ (b). For increasing field strength the eigenstates gain a finite admixture
of the quantum defect split states. For $600 \frac{V}{m}$ we clearly see a major contribution provided by the $38s$ state.}
\label{eigspec}
\end{minipage}
\end{figure}

In ref.\cite{Greene00} the authors reported on large electric dipole moments of ultralong-range polar Rydberg molecules of the order of
kDebye. The zero-field permanent dipole moment for these species scales according to the semi-classical
expression $D_{\rm{el}}=R_{\rm{eq}}-n^2/2$. In Fig.\ \ref{elec} we show the absolute value of the electric dipole moment
along the $z$-axis as a function of the field strength
\begin{eqnarray}
D_{\rm{el}}(E)&=&|\int \kd{d}^3\kd{r} \psi^{*}(\kd{r};\kd{R}_{\rm{eq}},\kd{E} ) z \psi(\kd{r};\kd{R}_{\rm{eq}},\kd{E} )| \nonumber\\
&=&\sqrt{\frac{4\pi}{3}}|\sum_{nn'll'm}C^{*}_{n'l'm}C^{*}_{nlm}\int dr r^3 R_{n'l'}(r)R_{nl}(r) \nonumber\\
&\times&\int d\Omega Y_{10}(\vartheta,\varphi)Y^{*}_{l'm}(\vartheta,\varphi)Y_{lm}(\vartheta,\varphi)|.\label{delint}
\end{eqnarray}  
The integration over the angular degrees of freedom provides $\Delta l=\pm1$ as a selection rule.
We observe that with increasing electric field also $D_{\rm{el}}$ increases up
to values of around $4$kDebye. As for $R_{\rm{eq}}$
we see a sharp step structure, i.e.\ for field strengths at approximately $100,\ 200,\ 385 \frac{V}{m}$ its values suddenly
increase in steps of roughly $500$ Debye. In Fig.\ \ref{elec} we also show a comparison between the exact result
calculated according to eq.\ (\ref{delint}) (blue data points) and the semi-classical approximation (green data points). For low electric fields the agreement is quite well, but differs with increasing field strength up to a deviation of around $10$\%. The semiclassical result therefore certainly allows for a qualitative description of the behavior of $D_{\rm{el}}$.
For $E> 570 \frac{V}{m}$ we find an unexpected decrease of $D_{\rm{el}}$. This feature can be understood if one analyzes the
field-dependent spectrum of coefficients for the electronic eigenvector
$\psi(\kd{r};\kd{R}_{\rm{eq}},\kd{E})=\sum_{i}C_{i}(E)\chi_{i}(\kd{r})$. In Fig.\ \ref{eigspec} we show the distribution
$|C_{i}|^2$ for $E=300 \frac{V}{m}$ and $600 \frac{V}{m}$. For $E=300 \frac{V}{m}$ the spectrum is dominated by basis states from the $n=35,\ l \ge 3$
manifold. Contributions stemming from the quantum defect split states (which are placed at the outermost right edge of the spectrum at $i=1217,...,1225$)
are negligible. For $E=600 \frac{V}{m}$ the situation has changed in the sense that now the main contribution is provided by the $38s$
state. This can be understood by the fact that the considered PES is approaching the $38s$ PES with increasing field strength.
The latter is however barely affected by the electric field. For $E=600 \frac{V}{m}$ the PES involve avoided crossings which causes the
high-$\ell$ dressed trilobite PES to acquire a major contribution from the $38s$ state. This finite admixture has two important consequences:\\
\begin{itemize}
\item Due to the $\Delta l=\pm 1$ selection rule the $38s$ state only acquires a contribution to the integral (\ref{delint})
via the $37p$ state. However, the coefficient of the latter state is negligibly small.
This causes the decrease of $D_{\rm{el}}$ for large field strengths as seen in Fig.\ \ref{elec}.\\ 
\item The finite $38s$ admixture provides us with the possibility to prepare high-$\ell$ Rydberg molecules via a two-photon process. 
This goes beyond the three-photon preparation scheme suggested in \cite{Greene00} ($l_{\rm{min}}=3$ for the field-free case).
For $E \ge 570 \frac{V}{m}$ the trilobite state acquires a major $l=0$ contribution which makes it accessible for
a two-photon transition scheme. The same mechanism has been reported recently in the analysis of ultralong-range polyatomic
Rydberg molecules formed by a polar perturber \cite{Rittenhouse11}. Field-free high-$\ell$ molecular states can then in
principle be accessed via an additional adiabatic switching of the electric field back to the zero value.
\end{itemize}
\section{Rovibrational states}
\begin{figure}[hbt]
\begin{minipage}{0.5\textwidth}
\includegraphics[width=1.0\linewidth]{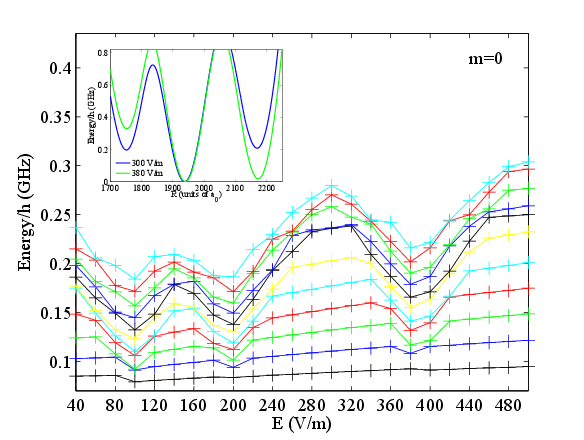} 
\caption{(Color online) Shown are the eleven lowest vibrational energies as a function of the field strength $E$. 
The dips around $E=100,\ 200$ and $385 \frac{V}{m}$ are caused by the change of potential wells
determining the diatomic equilibrium distance $R_{\rm{eq}}$. In the inset we show the offset corrected potential curves for $E=300\frac{V}{m}$ and $E=380\frac{V}{m}$ ($\theta=\pi$). For $E=300\frac{V}{m}$ bound states in the middle well with energies larger than $200$MHz can tunnel into the neighbored potential wells. This causes a reduction of their level spacings. For $E=380\frac{V}{m}$ we nearly get a double potential well and states with energies less than $200$MHz possess a higher tunneling probability. Correspondingly, this leads to a denser spectrum.}
\label{rovibra}
\end{minipage}
\end{figure}

\begin{figure}[hbt]
\begin{minipage}{0.5\textwidth}
\includegraphics[width=1.0\textwidth]{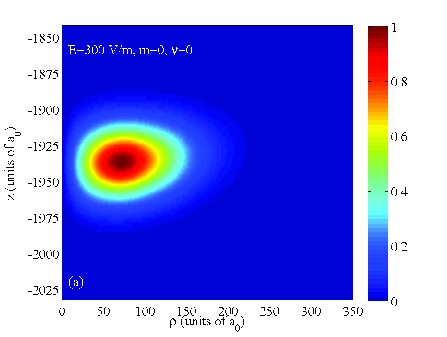}
\includegraphics[width=1.0\textwidth]{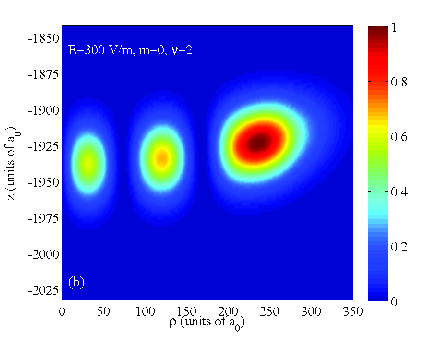}
\caption{(Color online) Scaled probability densities $|F_{k \nu m}(\rho,z)|^2$ for rovibrational wavefunctions. Both wavefuntions belong to the trilobite PES for $E=300 \frac{V}{m}$ with an azimuthal quantum number $m=0$. In (a) we observe a deformed Gaussian like density profile for the groundstate ($\nu=0$) centered at $Z=-1938a_0$ and $\rho=72a_0$. In $Z$/$\rho$-direction the density distribution has an extension of approximately $50a_0$/$100a_0$. In (b) we show the density profile for the second excited state ($\nu=2$). This density profile provided three peaks at $(Z,\rho)=(-1939a_0,32a_0),\ (-1934a_0,122a_0)$ and $(-1924a_0,234a_0)$.}
\label{probdensity}
\end{minipage} \end{figure}

Because of the azimuthal symmetry of the PES we introduce cylindric coordinates ($\rho,Z,\phi$)
for their parametrization $\epsilon(\kd{R})=\epsilon(\rho,Z)$. For the rovibrational wavefunctions we choose the following ansatz
\begin{eqnarray}
\hspace{-0.4cm}\phi_{k \nu m}(\kd{R})=\frac{F_{k \nu m}(\rho,Z)}{\sqrt{\rho}}\text{exp}(im \varphi),\ m \in \mathbb{Z},\ \nu \in \mathbb{N}_0 .
\end{eqnarray}
With this we can write the rovibrational Hamiltonian in eq.\ (\ref{hamrovi}) as 
\begin{eqnarray}
H_{\rm{rv}}&=&-\frac{1}{M}(\partial^{2}_{\rho}+\partial^2_{Z})+\frac{m^2-
1/4}{M\rho^2}+\epsilon(\rho,Z).
\end{eqnarray}
We solved the corresponding Schr\"odinger equation for different azimuthal quantum numbers $m$ using a fourth order finite
difference method.\\                                                                                                           
In Fig.\ \ref{rovibra} we provide the energies of the eleven lowest vibrational ($m=0$) states living in the trilobite PES for varying field strength. In order to obtain a normalized view of the spectrum the corresponding energy of the minimum of the PES has been subtracted. In general we observe a slight increase of the level spacing with increasing field strength. The increase is due to the an enhanced angular confinement of the rovibrational motion for strong fields. For $E=100,\ 200$ and $385\frac{V}{m}$ however we encounter a dip in the rovibrational level spacing. The latter corresponds to the case of crossover of the equilibrium positions between neighboring wells and therefore an accompanying relocation of the corresponding rovibrational wavefunctions. This leads to enhanced tunneling probabilities between neighboring wells and therefore an increased level density. In the inset of Fig.\ \ref{rovibra} we show the offset corrected potential curves for $E=300\frac{V}{m}$ (blue curve) and $E=380\frac{V}{m}$ (green curve) ($\theta=\pi$). For $E=300\frac{V}{m}$ the potential curve possesses a global minimum at $R=1939a_0$ and two local minima at $R=1750a_0$ and $ 2182a_0$ with an offset of $200$MHz. Bound states in the middle well with energies higher than $200$MHz can tunnel into these wells, whereby their level spacing is reduced. For increasing field strengths the right potential well is shifted downwards. This enhances the tunneling probabilities of states with energies less than $200$MHz, which correspondingly leads to a denser spectrum.
\\
In Fig.\ (\ref{probdensity}a,b) we present (scaled) probability densities $|F_{k \nu m}(\rho,z)|^2$ for $m=0$ for
the vibrational ground state ($\nu=0$) and the second excitation for $E=300 \frac{V}{m}$. 
The equilibrium distance for the PES is located at $Z=-1939a_0$, $\rho=0$. 
The $m=0$ ground state distribution is characterized by a deformed Gaussian profile that is localized at $Z=-1938a_0$ and $\rho=72a_0$. In $Z,\rho$-direction the density distribution possesses an extension of approximately $50a_0$ and $100a_0$, respectively.
The density profile for the second excitation ($\nu=2$) shows three separate Gaussian like density
peaks with increasing intensity located at $(Z,\rho)=(-1939a_0,32a_0),\ (-1934a_0,122a_0)$ and $(-1924a_0,234a_0)$ with an extension of around ($25a_0$, $30a_0$), ($25a_0,40a_0$) and ($75a_0,75a_0$) in the $Z,\rho$-directions, respectively. 
\section{Conclusions}
The recent spectacular experiments \cite{Bendkowsky09,Pfaunat11,Bendkowsky10} preparing, detecting and probing some of the important
properties of non-polar ultralong-range Rydberg molecules have opened the doorway towards a plethora of possibilities to create new
exotic species where atoms, molecules or even clusters and mesoscopic quantum objects might be bound to electronic Rydberg systems. 
It is therefore of crucial importance to learn how the properties of these Rydberg molecules can be tuned finally leading to a control
of the structure and potentially dynamics of these systems. The primary choice are here external fields in particular due to the susceptibility
of the weakly bound Rydberg electrons. In the present work we have therefore explored the changes the polar high angular momentum
trilobite states experience 
if they are exposed to an electric field of varying strength. Taking into account $s$- and $p$-wave interactions it turns out that
the electric field provides us with a unique know to control the topology of the adiabatic potential energy surface. First of all,
the angular degree of freedom between the electric field and internuclear axis is converted from a rotational to a vibrational degree
of freedom thereby rendering the field-free potential energy curve into a two-dimensional potential energy surface. It turns out that
the global equilibrium position is always the antiparallel configuration of these two axes. The sequence of potential wells with increasing 
radial coordinate, i.e.\ the oscillatory behavior of the potential is changed dramatically in the presence of the field. In particular 
we encounter an overall lowering of the energy accompanied by a subsequently crossover of the energetically order of the individual wells.
Consequently, the equilibrium distance and the lowest vibrational states are systematically shifted to larger internuclear distances.
The $p$-wave split state which, due to its resonant behavior, lowers dramatically in energy with decreasing internuclear distance
and therefore crosses the polar trilobite state close to its equilibrium distance in the zero-field case, can now with increasing
field strength be systematically shifted away from the energetically lowering equilibrium distance and corresponding well.
In such a way the respective stability of the ground and many excited vibrational states of the polar trilobite state is guaranteed.
For strong fields the interaction of the latter state with non-polar (quantum defect split) states, which are very weakly polarized
in the presence of the field, leads to a strong admixture of, in our specific case, $s$-wave character to the polar high angular 
momentum states. As a consequence, a two-photon excitation process starting from the ground state of the two-atom system
should be sufficient to efficiently excite these states and probe their character. The electric dipole moment, which is steadily
increasing with increasing electric field strength starting from zero-field, does, due to the above admixture, decrease in the
strong field regime.
\\ 
To obtain an even richer topology of the potential energy surfaces of the trilobite states the combination of static electric and
magnetic field would be the next step. In this case both rotational degrees of freedom in field-free space will, in general, turn
into vibrational modes rendering the field-free potential energy curve a three-dimensional potential energy surface. Depending on the
configuration, such as parallel or crossed fields, the remaining symmetries might even lead to controllable crossings or avoided
crossings of the surfaces which will be the subject of a future investigation.
\section{Acknowledgment}
We thank the Initial Training Network COHERENCE of the European Union FP7 framework for financial support. In addition, we thank Michael Mayle and Igor Lesanovsky 
for helpful discussions and suggestions. One of the authors (P.S.) acknowledges the hospitality and many
fruitful discussions in particular with H.R. Sadeghpour at the Institute for Theoretical Atomic Molecular
and Optical Physics at the Harvard Smithsonian Center for Astrophysics in Cambridge, USA.  
\end{document}